\documentclass[prl,twocolumn,superscriptaddress,floatfix]{revtex4}
\usepackage{amsmath,amsfonts,amssymb,graphics,graphicx,epsfig,color,natbib}

\newcommand{\be}{\begin{equation}}
\newcommand{\ee}{\end{equation}}
\newcommand{\ba}{\begin{eqnarray}}
\newcommand{\ea}{\end{eqnarray}}
\newcommand{\ban}{\begin{eqnarray*}}
\newcommand{\ean}{\end{eqnarray*}}

\newcommand{\ket}[1]{\mbox{$ | #1 \rangle $}}

\newcommand{\si}{\sigma}

\newcommand{\compl}{\begin{picture}(8,8)\put(0,0){C}\put(3,0.3){\line(0,1){7}}\end{picture}}
\newcommand{\real}{\begin{picture}(8,8)\put(0,0){R}\put(0,0){\line(0,1){7}}\end{picture}}

\newcommand{\one}{\leavevmode\hbox{\small1\normalsize\kern-.33em1}}

\begin{document}

\title{Testing the Hilbert space dimension}
\author{Nicolas Brunner}
\address{Group of Applied Physics, University of Geneva, Geneva, Switzerland}
\author{Stefano Pironio}
\address{ICFO-Institut de Ciencies Fotoniques, Castelldefels (Barcelona), Spain}
\author{Antonio Acin}
\address{ICFO-Institut de Ciencies Fotoniques, Castelldefels (Barcelona), Spain}
\address{ICREA-Instituci\'o
Catalana de Recerca i Estudis Avan\c cats, Barcelona, Spain}
\author{Nicolas Gisin}
\author{{Andr\'e Allan M\'ethot}}
\address{Group of Applied Physics, University of Geneva, Geneva, Switzerland}
\author{Valerio Scarani}
\address{Centre for Quantum Technologies, National University of Singapore, Singapore}
\date{\today}

\begin{abstract}
Given a set of correlations originating from measurements on a
quantum state of unknown Hilbert space dimension, what is the
minimal dimension $d$ necessary to describes such correlations? We
introduce the concept of dimension witness to put lower bounds on
$d$. This work represents a first step in a broader research program
aiming to characterize Hilbert space dimension in various contexts
related to fundamental questions and Quantum Information
applications.
\end{abstract} \maketitle

A theorist is invited to visit a lab. The experimentalists, not
entirely happy with the nuisance, decide to submit the visitor to
the ordeal ``Guess what we are measuring''. Hardly distinguishing
lasers from vacuum chambers, the theorist cannot hope to identify
the system under study, and asks for a black-box description of
the experiment in order to disentangle at least the physics from
the cables. It turns out that the experiment aims at measuring
correlations between the outputs of two measuring apparatuses. On
each side, the outcome of the measurement is discrete and can take
$v$ values --- the theorist writes $a,b\in\{0,1,...,v-1\}$. A knob
with $m$ positions allows to change the parameters of each
measuring apparatus -- the theorist writes
$x,y\in\{0,1,...,m-1\}$. Finally, the experimentalists show the
data: the frequencies $P(ab|xy)$ of occurrence of a given pair of
outcomes for each pair of measurements. The theorist makes some
calculations and delivers a verdict...

Some verdicts have been known for some time. In particular, if
$P(ab|xy)$ violates a Bell-type inequality \cite{Bell64}, we know
for sure that an entangled quantum state has been produced in the
lab. If on the contrary $P(ab|xy)$ can be distributed by shared
randomness, the experiment may in fact be purely classical.

The goal of this paper is to introduce another family of verdicts,
different from the ``quantum-vs-classical'' one. We prove that,
even in a black-box scenario, the theorist may have something to
say about the \textit{dimension of the Hilbert space} of the
quantum objects that are measured. Both the enthusiastic verdict
``You are using systems of dimension at least $d$'' and the
disappointing one ``You may be coding in less than $d$
dimensions'' are possible.

From an information-theoretical point of view, the dimensionality
of quantum systems can be seen as a resource. Thus testing the
Hilbert space dimension is important for quantifying the power of
quantum correlations, a central issue in Quantum Information
science. Furthermore, this line of research turns out to be
relevant for Quantum Key Distribution (QKD) as well. In standard
security proofs of QKD \cite{shorpreskill}, the correlations
shared by the authorized partners, Alice and Bob, are supposed to
come from measurements on a quantum state of a given dimension.
This assumption turns out to be crucial for the security of most
of the existing protocols \cite{NLcrypto}. But is the dimension of
a quantum system an experimentally measurable quantity? There also
exist protocols whose security does not require any hypothesis on
the Hilbert space dimension \cite{DevIndep}. However, to prove
security in such protocol, it is useful to understand how it is
possible to bound effectively the dimension of the systems
distributed by the eavesdropper \cite{mayersyao}.

Formally, a set of conditional probabilities $P(ab|xy)$ has a
$d$-dimensional representation if it can be written as
\begin{equation}\label{quditpr}
P(ab|xy)=\text{tr}(\rho M^X_a\otimes M^Y_b) ,
\end{equation}
for some state $\rho$ in $\compl^d\otimes\compl^d$ and local
measurements operators $M^X_a$ and $M^Y_b$ acting on $\compl^d$,
or if it can be written as a convex combination of probabilities
of the form \eqref{quditpr}. We are interested in the following
question: what is the minimal dimension $d$ necessary to reproduce
a given set of probabilities $P(ab|xy)$ \cite{noted}?

The fact that we allow convex combinations of \eqref{quditpr}
means that shared randomness is unrestricted in our scenario. This
is consistent with a quantum information perspective where
classical resources are taken to be free and we want to bound the
quantum resources, in this case the dimensionality of the quantum
states, necessary to achieve a task. Within this approach, the
answer to the above question is immediate if the initial
correlations admit a locally causal model \cite{Bell64}, as in
this case they can be reproduced using shared randomness only and
no quantum systems are strictly needed for their preparation.
Thus, our problem is interesting only when the initial
correlations are non-local.

Since classical correlations are taken to be free, the set of
$d$-dimensional quantum correlations is convex. Therefore,
standard techniques from convex theory can be applied, as has been
done for other quantum information problems such as separability
\cite{EntWit}. Following this analogy, we introduce the concept of
dimension witnesses. A $d$-dimensional witness is a linear
function of the probabilities $P(ab|xy)$ described by a vector
$\vec w$ of real coefficients $w_{abxy}$, such that
\begin{equation}\label{dimwitn}
    \vec w\cdot\vec p \equiv \sum_{a,b,x,y} w_{abxy}P(ab|xy)\leq
    w_d
\end{equation}
for all probabilities of the form \eqref{quditpr} with $\rho$ in
$\compl^d\otimes\compl^d$, and such that there are quantum
correlations for which $\vec w\cdot\vec p>w_d$. When some
correlations violate \eqref{dimwitn}, they can thus only be
established by measuring systems of dimension larger than $d$.
Dimension witnesses allow us to turn the Hilbert space dimension,
a very abstract concept, into an experimentally measurable
property.

In the following, we construct several examples of 2-dimensional
witnesses. We also show that not all $2$-outcome quantum
correlations are achievable with qubits, answering a question
raised by Gill \cite{gill}. A proof of the same result for two
parties has been independently obtained in \cite{tamas}, while the
results of \cite{perezgarcia} answer Gill's question in the
tripartite case.

\paragraph{Witnesses based on CGLMP.}A natural starting point for our investigations is the situation
corresponding to $m=2$ measurement settings per side with $v=3$
possible outcomes. Indeed, in this case
Collins-Gisin-Linden-Massar-Popescu (CGLMP) introduced a Bell
inequality whose maximal quantum violation is achieved by a
two-qutrit state. The CGLMP expression is
\be\label{CGLMP}\begin{split}
C(\vec p)&= P(b_0\geq a_0)+P(a_0\geq b_1)+P(a_1\geq b_0)
\\   &\quad + P(b_1>a_1)-3
\end{split}\ee
where $P(a_x\geq b_y)=\sum_{a\geq b}P(ab|xy)$ \cite{CGLMP,gill2}.
Local correlations satisfy $C(\vec p)\leq 0$, while measurements on
a partly entangled two-qutrit state yields a maximal value of
$C(\vec p)=0.3050$ \cite{toni}.

The set of quantum probabilities corresponding to $m=2$ and $v=3$
lives in a 24-dimensional space. Since it is in general difficult
to gain intuition in such a high-dimensional space, we will focus
here on a two-dimensional subspace of this quantum set, which has
been characterized in \cite{praNScrypto}. This subspace is
parameterized by two numbers: the CGLMP value $C(\vec p)$ and
\begin{equation*}D(\vec
p)=-\sum_{x,y=0}^1\sum_{k=0}^2P(a=k,b=k-1-(x-1)(y-1)|xy)
\end{equation*}
The precise form of the probabilities living in this subspace
as a function of $C(\vec p)$ and $D(\vec p)$ can be found in
\cite{praNScrypto}. Note that if two parties share a point $\vec p$
in the original quantum set, they can run a depolarization protocol
that will map it onto the two-dimensional subspace while keeping the
values of $C(\vec p)$ and $D(\vec p)$ constant \cite{praNScrypto}.

Since the quantum region is convex, its boundary in the
two-dimensional subspace can be obtained by computing the maximal
value of \be\label{cd} I_\phi(\vec p)=\cos\phi C(\vec p) +\sin\phi
D(\vec p) \ee for all $\phi$, that is by computing how far it
extends in every direction of the two-dimensional subspace. We
have computed these values using the technique introduced in
\cite{miguel}. The optimal values $I_{\phi}^{(q)}$ are obtained
for entangled states of the form
$\ket{\psi}=\frac{1}{\sqrt{2+\gamma^2}}(\ket{00}+\gamma
\ket{11}+\ket{22})$ with measurements that are independent of
$\gamma$ and which can be found in \cite{CGLMP}. The resulting
quantum curve is represented on Fig.~1 using the parametrization
of \cite{praNScrypto}.
\begin{figure}[t]
\begin{center}
\includegraphics[width=0.75\columnwidth]{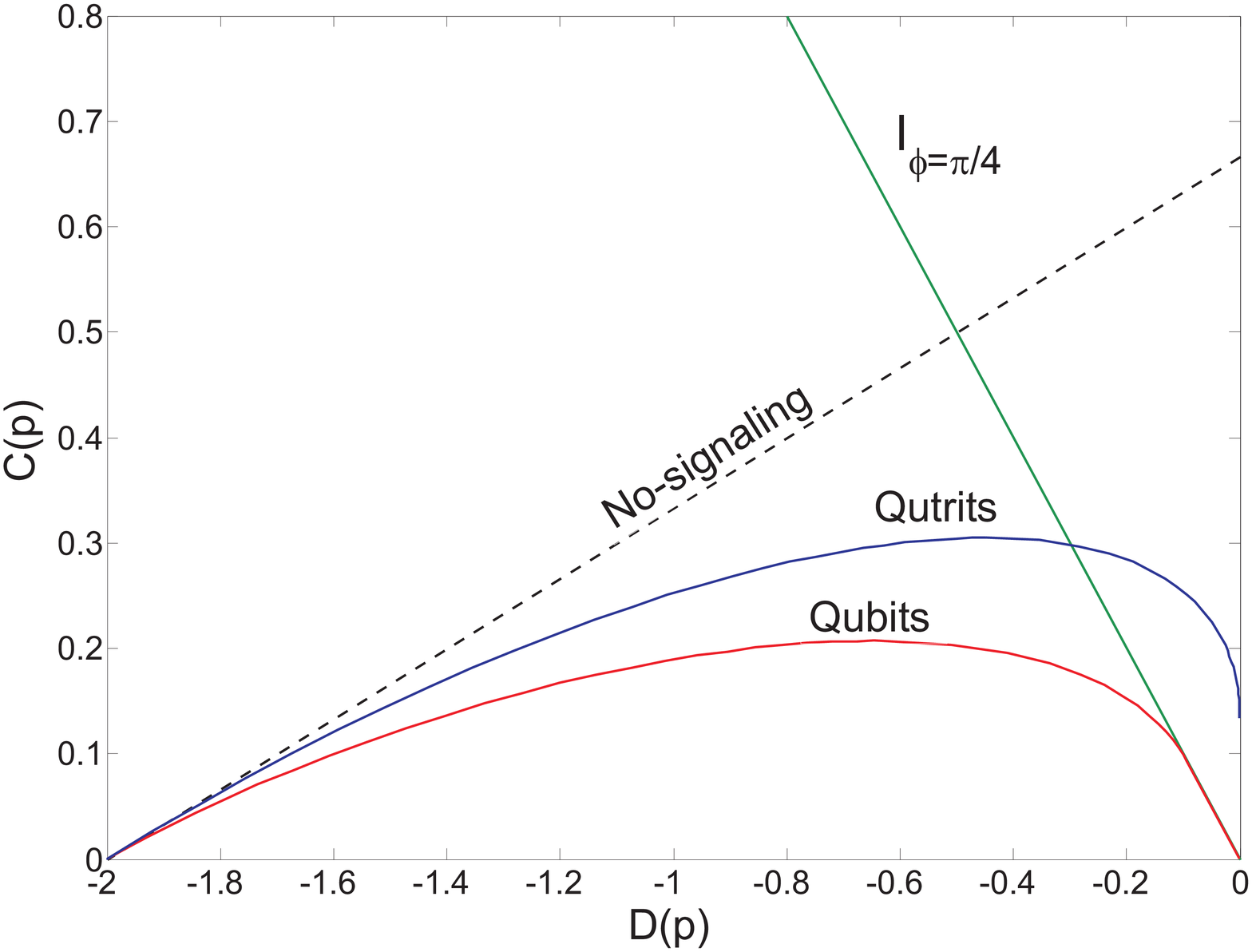}
\caption{Quantum region in the two-dimensional subspace described
in \cite{praNScrypto}. The upper curve represents the boundary of
the general quantum region and can be achieved by measurements on
two-qutrit states. The lower curve represents the boundary of the
region accessible through measurements on two-qubit states. The
dashed line delimits the no-signaling correlations. The inequality
$I_{\phi=\pi/4}\leq 0$ is a dimension witness: it cannot be
violated by performing measurements on qubits, but qutrits are
required.} \label{Triangle}
\end{center}
\end{figure}

We have also determined the region accessible with qubits by
maximizing $I_\phi$ over all measurements (POVMs), and two-qubit
states. The optimal values $I_{\phi}^{(2)}$ are obtained by
performing two-outcome von Neumann measurements on pure entangled
two-qubit states
$\ket{\psi}=\cos{\theta}\ket{00}+\sin{\theta}\ket{11}$. The
resulting curve is also shown in Fig.~1. Note that contrarily to
the previous case the qubit curve is not the result of an exact
computation, but of a numerical search using a heuristic
algorithm. Indeed, the method of \cite{miguel} cannot be directly
applied here, as it does not constrain the dimension of the
quantum systems, while others techniques based on semidefinite
programming \cite{lasserre} are computationally too costly.


The inequalities $I_\phi(\vec p)\leq I_\phi^{(2)}$ form a family
of 2-dimensional witnesses. Any one of these inequalities for
which the maximal quantum value $I_\phi^{(q)}>I_\phi^{(2)}$ is
strictly greater than the maximal qubit value, that is any
direction in Fig.~1 for which there is a gap between the qubit and
the general quantum curve, allows one to distinguish qubits from
higher-dimensional systems. Note that the expressions \eqref{cd}
can also be interpreted as Bell inequalities with local bound
$I_\phi^{(l)}=0$ if $\sin\phi$ is positive and
$I_\phi^{(l)}=-2\sin(\phi)$ otherwise. The inequalities with
$\tan{\phi}\geq 1$ are noteworthy because the local bound and the
qubit bound coincide, $I^{(2)}_\phi=I^{(l)}_\phi=0$, i.e., qubits
no longer violate them; they can only be violated with qutrits or
higher dimensional systems.

Although the situation that we just considered is illustrative, we
mentioned that we had to resort to heuristic numerical searches to
compute the qubit value of the expressions \eqref{cd}. We now
present two situations were stronger statements can be made. While
techniques have been developed to characterize the boundary of the
general quantum set (i.e., with no bound on the dimension)
\cite{miguel}, we still lack of efficient tools to characterize the
quantum region corresponding to fixed Hilbert-space dimension. The
two examples below provide two different approaches to this problem,
the first one uses semidefinite programming, the second one
establishes a link with the Grothendieck constant.

\textit{Using semidefinite programming.} We give here an example of
a dimension witness where the maximal violation can be determined
for any two-qubit state using semidefinite programming
\cite{semidefinite}. We consider a scenario where Alice chooses
between two settings ($m_A=2$) and Bob among three settings
($m_B=3$). All settings yield binary outcomes except Alice's second
setting $x=1$, which is ternary. In this case, the following Bell
expression
\be\label{Istef}
\begin{split}E(\vec p)&=P_A(0|0)-P(00|00)-P(00|01)-P(00|02)\\
&\quad+P(00|10)+P(10|11)+P(20|12)+1\end{split}\ee with local bound
$E(\vec p)\leq 0$, has recently been introduced \cite{Allchsh}.
The maximal quantum violation $E^q=0.2532$ can be found using the
method of \cite{miguel} and is achieved for a partially entangled
state of two-qutrits.

In order to prove that the largest violation for qubits is
strictly smaller, we computed the maximal value of the right-hand
side of \eqref{Istef} over all two-qubit states
$\rho\in\mathbb{C}^2\otimes\mathbb{C}^2$ and over all measurement
settings. Since we seek to maximize an expression which is linear
in the probabilities $p(ab|xy)=\text{tr}\left[\rho M^x_a\otimes
M^y_b\right]$, its maximum will be attained by \emph{pure} states
$\rho=\ket{\psi}\langle\psi|$ and \emph{extremal} POVMs. Up to a
local change of basis, any pure two-qubit state can be written as
$\ket{\psi(\theta)}=\cos(\theta)\ket{00}+\sin(\theta)\ket{11}$.
Every extremal POVM $M$ for qubits has elements $\{M_i\}$ which
are proportional to rank $1$ projectors \cite{dariano} and can
thus be parameterized in term of the Pauli matrices
$\vec{\sigma}=\left(\si_x,\si_y,\si_z\right)$ as $
M_i=\frac{1}{2}\left( m_i \one +\vec{m}_i\cdot\vec{\sigma}\right)$
where $m_i\geq 0$, $\sum_i m_i=2$, $\sum_i \vec{m}_i=0$, and
$(m_i)^2=(\vec{m}_i)^2$. Let $u$ denote the set of variables
necessary to represent all POVMs using this parameterizations, and
let $c(u)\geq 0$ represent the (quadratic) constraints to which
these variables are subject. For given $\theta$, the right-hand
side of \eqref{Istef} is a quadratic function $E_\theta(u)$ of
$u$. Our problem is thus to solve the following (non-convex)
quadratic program
\begin{equation}\label{optim}
E_\theta^\ast = \max_u\; E_\theta(u)\quad \text{s.t.}\quad c(u)\geq
0 \, .
\end{equation}
Solving such a problem is in general a difficult task, as it may
have many local optima. Following the approach of Lasserre
\cite{lasserre} we derived upper-bounds on $E_\theta^\ast$ using
semidefinite programming \cite{semidefinite}. For any given value
of $\theta$, we obtained an upper-bound on the maximal value of
the right-hand side of \eqref{Istef}. This value coincides up to
numerical precision with the maximum value obtained when we
discard one of the outcomes of the POVM $x=1$. In this case the
inequality \eqref{Istef} reduces to the CHSH inequality
\cite{chsh}, whose maximal violation as a function of $\theta$ is
$(\sqrt{1+\sin^2(2\theta)}-1)/2$. The maximal qubit violation of
\eqref{Istef} is thus equal to $E^{(2)}=1/\sqrt{2}-1/2 \approx
0.2071$, to be compared with the maximal quantum violation
$E^{(q)}=0.2532$ achieved using two-qutrit states. Let us stress
that our qubit bound, which can be reached, is an upper-bound on
the global optimal solution of the problem, since there exist
algorithms able to find the \emph{global} optimum of semidefinite
programs \cite{semidefinite}.

\textit{Link to the Grothendiek constant.} The previous examples
of qubit witnesses all contain at least one three-outcome
measurement. In this case, it is perhaps not surprising, though
difficult to prove, that systems of dimension larger than two are
needed to get the maximal quantum value. In what follows, we show
that qubit witnesses exist even for two-outcome correlations,
answering Gill's question \cite{gill,tamas,perezgarcia} in the
bipartite case.

Define the correlator $c_{xy}$ between measurement $x$ by Alice
and $y$ by Bob as $c_{xy}=P(a=b|xy)-P(a\neq b|xy)$, and consider
now a linear function of such correlators,
\begin{equation}\label{gencomb} I=\sum_{i,j=1}^m M_{ij}
c_{x_iy_j}\end{equation} defined by an $m\times m$ matrix $M$
verifying the normalization condition
$\max_{\{x_i,y_j\}}|\sum_{i,j=1}^m M_{ij}\,x_iy_j|=1$ with
$x_i,y_j=\pm 1$. Because of this normalization, $I$ can be seen as
a standard Bell inequality with local bound 1.

On the other hand, the correlators are quantum, i.e.
$c_{xy}=\langle X\otimes Y\rangle_\psi$ for some observables $X$
and $Y$ with $\pm 1$ eigenvalues, if and only if there exist two
normalized vectors $\vec x,\vec y \in \real^N$ such that
$c_{xy}=\vec x\cdot \vec y$ (see \cite{tsirelson3,AGT} for
details). The maximum value that any operator $I$ can take, when
$c_{xy}$ is of this form, is known in the mathematical literature
as $K_G(N)$ and called the \textit{Grothendieck constant of order
$N$}; the maximum over all $N$ is written $K_G$. Note now that in
the case of two-outcome correlators the analysis can be restricted
to projective measurements \cite{ww}. In this situation, any Bell
operator associated to $I$ is diagonal in the Bell basis, implying
that the largest value is obtained for a maximally entangled
state, say the singlet. Since any two-outcome correlator for
projective measurements on the singlet state is equal to the
scalar product of three-dimensional real vectors, the maximal
value of $I$ achievable with qubits is $K_G(3)$. Although the
exact values of the Grothendieck constants are still unknown, it
is proven that $K_G(3)<K_G$ \cite{footnote}: this means that there
exist an inequality $I$ which is not saturated by correlations
coming from two qubits. This proves the existence of dimension
witnesses for qubits with two-outcome measurements. Examples of qubit witnesses built from
two-outcome measurements were recently found in \cite{tamas}.

We conjecture that two-outcome measurements may be sufficient to
test the dimension of any bipartite quantum system, in the sense
that there exist dimension witnesses built from binary
measurements for any finite dimension. Indeed, all quantum
correlators in $\compl^d\otimes\compl^d$ can be written as scalar
product of vectors of size $2d^2$ \cite{tsirelson3,AGT}.
Therefore, if $K_G(N)$ is strictly smaller than $K_G$ for any
finite $N$, which is plausible but unproven to our knowledge, one
can construct witnesses with binary measurements for arbitrary
dimension.

\paragraph{Conclusion and other directions.}
With the goal of testing the Hilbert space dimension of an unknown
quantum system, we introduced the concept of dimension witness. We
presented two examples of qubit witnesses, which can detect
correlations that require measurements on quantum systems of
dimension greater than two for their generation. Both of these
examples involved three-outcome measurements; so the number of
measurement outcomes exceeded the dimension of the Hilbert space
to be witnessed. This shows, as one may expect, that not all
$d$-outcome correlations can be obtained by measuring quantum
systems of dimension smaller than $d$. Then, somehow surprisingly,
we proved that qubit witnesses also exist in the case of
two-outcome measurements.

Viewing the dimensionality of a quantum system as a resource and
trying to understand how to estimate or bound it, is an approach
that deserve further investigation. The concept of dimension
witnesses represents only a first step in this direction. In
another direction, it was shown in \cite{mayersyao} how it is
possible to bound, even in an adversary scenario, the dimension
necessary to represent a given set of correlations.

In general, the problem of Hilbert space characterization is not
restricted to a multipartite non-local scenario. When dealing with
fundamental issues for instance, it can be relevant to estimate
the dimension of a quantum system without any distinction between
classical and quantum resources. The motivation being that if
nature is indeed described by quantum theory, classical degrees of
freedom have also to be coded ultimately in quantum systems. In
this context the (possibly one-partite) global quantum state is
then taken to be pure and the goal is, given some initial
statistical data, to determine the physical realization of minimal
dimension. One may also wonder to what extent high dimensional
quantum systems are more powerful than lower dimensional ones when
noise is added. In particular, it would be interesting to look for
dimension witnesses very robust to noise. Finally, a related,
though different question concerns the multi-partite case. How can
one be sure that data obtained by measurements on a $n$-party
quantum state do require $n$-partite entanglement without any
assumption of the local Hilbert space dimensions?

The authors thank C. Branciard, D. P\'erez-Garc\'{\i}a, S.
Popescu, and T. V\'ertesi for useful discussions and acknowledge
financial support from the EU project QAP (IST-FET FP6-015848),
Swiss NCCR Quantum Photonics, Spanish MEC through Consolider QOIT
and FIS2007-60182 projects, a ``Juan de la Cierva" grant, and the
National Research Foundation and Ministry of Education, Singapore

\bibliographystyle{prsty}
\bibliography{DimH_7}

\begin{thebibliography}{10}
\bibitem{noted}
{O}ne can always increase the dimension of the Hilbert space by adding useless
  extra dimensions. Thus, the relevant question here is to obtain the minimal
  dimension $d$ required for the given correlations.
\end{thebibliography}

\end{document}